\documentclass[%
reprint,
superscriptaddress,
showpacs,
aps,
prb,
]{revtex4-1}

\usepackage[version=4]{mhchem}
\usepackage[separate-uncertainty = true,multi-part-units=single]{siunitx}
\usepackage{amsmath,amssymb}
\usepackage{multirow}
\usepackage{graphicx}
\usepackage{amsbsy}
\usepackage{color}
\usepackage{tabularx, booktabs}
\usepackage[colorinlistoftodos]{todonotes}
\usepackage{pbox}
\usepackage{braket}
\usepackage[caption=false,position=b,singlelinecheck=off,font=normalsize,labelfont=bf,justification=justified]{subfig}
\usepackage{xr}
\usepackage{hyperref}
\hypersetup{
	colorlinks=true,
	linkcolor=blue,
	filecolor=black,      
	urlcolor=blue,                                                                       
	citecolor=blue,}

\usepackage[capitalise,noabbrev,nameinlink]{cleveref}  

\begin{document}
	
	\title{Revealing the site selective local electronic structure of Co$_{3}$O$_{4}$ using $2p3d$ resonant inelastic X-ray scattering}
	
	\author{Ru-Pan Wang}
	\affiliation{Debye Institute for Nanomaterials Science, Utrecht University, Universiteitsweg 99, 3584 CG Utrecht, The Netherlands}
	\affiliation{Department of Physics, University of Hamburg, Luruper Chaussee 149, G600, 22761 Hamburg, Germany}
	\affiliation{Deutsches Elektronen-Synchrotron DESY, Notkestraße 85, 22607 Hamburg, Germany}
	
	\author{Meng-Jie Huang}
	\affiliation{Deutsches Elektronen-Synchrotron DESY, Notkestraße 85, 22607 Hamburg, Germany}
	\affiliation{Karlsruhe Institute of Technology (KIT), Institute for Quantum Materials and Technologies (IQMT), 76021 Karlsruhe, Germany}
	
	\author{Atsushi Hariki}
	\affiliation{Department of Physics and Electronics, Graduate School of Engineering, Osaka Prefecture University 1-1 Gakuen-cho, Nakaku, Sakai, Osaka 599-8531, Japan}
	
	\author{Jun Okamoto}
	\author{Hsiao-Yu Huang}
	\author{Amol Singh}
	\author{Di-Jing Huang}
	\affiliation{National Synchrotron Radiation Research Center, No.101 Hsin-Ann Road, Hsinchu Science Park, Hsinchu 30076, Taiwan}
	
	\author{Peter Nagel}
	\author{Stefan Schuppler}
	\affiliation{Karlsruhe Institute of Technology (KIT), Institute for Quantum Materials and Technologies (IQMT), 76021 Karlsruhe, Germany}
	
	\author{Ties Haarman}
	\affiliation{Debye Institute for Nanomaterials Science, Utrecht University, Universiteitsweg 99, 3584 CG Utrecht, The Netherlands}
	
	\author{Boyang Liu}
	\thanks{boyang\_liu@vip.sina.com}
	\affiliation{Debye Institute for Nanomaterials Science, Utrecht University, Universiteitsweg 99, 3584 CG Utrecht, The Netherlands}
	
	\author{Frank M. F. de Groot}
	\thanks{F.M.F.deGroot@uu.nl} 
	\affiliation{Debye Institute for Nanomaterials Science, Utrecht University, Universiteitsweg 99, 3584 CG Utrecht, The Netherlands}
	
	
	\begin{abstract}
		We investigate mixed-valence oxide Co$_3$O$_4$ using Co $2p3d$ resonant inelastic X-ray scattering (RIXS). 
		By setting resonant edges at Co$^{2+}$ and Co$^{3+}$ ions, the $dd$ excitations on the two Co sites are probed selectively, providing detailed information on the local electronic structure of Co$_3$O$_4$.
		The $2p3d$ RIXS result reveals the $^4$T$_{2}$ excited state of tetrahedral Co$^{2+}$ site at 0.5 eV beyond the discriminative power of optical absorption spectroscopies. 
		Additionally, the $^3$T$_{2g}$ excited stated at 1.3 eV is uniquely identified for the octahedral Co$^{3+}$ site. 
		Guided by cluster multiplet simulations, the ground-state character of the Co$^{2+}$ and Co$^{3+}$ site is determined to be high-spin $^4$A$_{2}$(T$_d$) and low-spin $^1$A$_{1g}$(O$_h$), respectively.
		This indicates that only the Co$^{2+}$ site is magnetically active site at low-temperatures in Co$_3$O$_4$. 
		The ligand-to-metal charge transfer analysis suggests a formation of a strong covalent bonding between Co and O ions at the Co$^{3+}$ site, while Co$^{2+}$ is rather ionic.
		\\
		
	\end{abstract}
	
	\maketitle
	
	\section{Introduction}
	Electronic properties in materials with strongly interacting electrons, such as $3d$ transition-metal (TM) oxides, are governed by an inter-atomic hybridization (covalent bonding) building on the multiplet spectrum that is determined by Hund's coupling, crystal-field splitting and spin-orbit coupling. 
	The determination of the local electronic structure is a relatively straightforward task for localized materials, but it faces an experimental challenge for a mixed-valence TM compound, especially when the constituting TM ions have rich overlapping multiplet structure, which is the case for Co$_3$O$_4$.
	
	Co$_3$O$_4$ crystalizes in a normal spinel structure (AB$_2$O$_4$) below 850 K~\cite{Picard1980jlcm, Sparks2018prb}.
	The Co ions in the A site with tetrahedral (T$_d$) local environment are divalent (Co$^{2+}$), while those in the B sites with octahedral (O$_h$)\footnote{The crystal structure analysis indicated that the Co$^{3+}$ ions are located in a trigonal (pseudo O$_h$) local environment. However, our discussions follow the generally indication of O$_h$ symmetry at the B site.} local environment are trivalent (Co$^{3+}$). 
	An experimental determination on the multiplet character of two inequivalent Co sites is of fundamental importance to elucidate the magnetic coupling responsible for the antiferromagnetic order (below $\sim$40~K)~\cite{Roth1964jpcs, Mironova1994ssc}. 
	The Co-site dependence of covalency is crucial for large exchange anisotropies achieved by a substitution of Ni or Fe for Co ions and its application to magneto-optical information storage~\cite{Goodenough1955pr, Martens1985jpcs, Vaz2010prb}. 
	The covalency also influences charge capacity.
	Co$_3$O$_4$ shows a high charge capacity (700~mAh/g) and a good cycle performance on the nano-size negative-electrodes, which elevates Co$_3$O$_4$ to an influential compound for the lithium battery~\cite{Poizot2000nature, Du2007am}.
	The catalytic activity of the Co$^{2+}$ ions in Co$_3$O$_4$ is actively debated~\cite{Kim2014jpcl, Wang2016jacs}; it is important for the oxygen evolution reaction step in the application of water oxidation~\cite{Li2014csr, Singh2007ec, Esswein2009jpcc, Jiao2009acie, Xi2012jpcc, Zhang2018jpcc}. 
	The determination of the optical band gap is also relevant to the local electronic structure since a small band gap of Co$_3$O$_4$ in the visible region was found and allows the application in photovoltaic cells~\cite{Singh2015jctc, Qiao2013jmcc}. 
	
	The electronic structure of Co$_3$O$_4$ has been characterized by the
	ultraviolet/visible (UV/Vis)~\cite{Belova1983ssc, Cook1986tsf, Miedzinska1987jpcs, Wang2004jpcb, Lima2014jpcs} and near infrared (NIR) absorption~\cite{Wood1967jpc, Mironova1994ssc}. 
	These absorption spectra show excitations at 0.82, 0.93, 1.64, and 2.81 eV~\cite{Cook1986tsf}, while diverse interpretations for these excitations are proposed~\cite{Kim2014jpcl, Wang2004jpcb, Belova1983ssc, Miedzinska1987jpcs, Mironova1994ssc, Lima2014jpcs},
	partly because the direct $dd$ excitations are dipole forbidden that limits their discriminative power~\cite{vanSchooneveld2013acie, Liu2016ic}.
	The tetrahedral symmetry of the Co$^{2+}$ site effectively allows $dd$ excitations, but the excitations lower than 0.82~eV are still an open question~\cite{Mironova1994ssc, Wood1967jpc}. 
	Hibberd \textit{et al.} performed Co $L_{2,3}$-edge ($2p$) x-ray absorption spectroscopy (XAS)~\cite{Hibberd2015jpcc}. 
	With the aid of Co-based polyoxometalates (POMs), they revealed distinct absorption peaks for the Co valence (2$+$ or 3$+$) and local environment (O$_h$ or T$_d$) in Co $L_3$-XAS spectra. 
	However, due to possible overlap of the multiple Co-site signals and the lifetime broadening ($\sim$200~meV), the reliability of the information on the local electronic structure is limited. 
	
	In this article, we study the Co-site resolved low-energy local excitations in Co$_3$O$_4$ using Co $2p3d$ resonant inelastic x-ray scattering (RIXS). 
	By setting the resonant photon energy to the distinct Co $L_3$-XAS features of the Co$^{2+}$ and Co$^{3+}$ sites \cite{Hibberd2015jpcc}, the site-selective local excitations are measured,
	which is referred to as site-resolved RIXS in recent literature~\cite{vanSchooneveld2013acie, Elnaggar2020prb, winder2020xray, Lu2018prx}.
	The (\textit{photon-in--photon-out}) transitions in RIXS brings a broad sensitivity to $dd$ excitations of the system, which provides a better determination on the local electronic structure including the orbital hybridization effects~\cite{Magnuson2002prb, Wang2017jpcc, Hariki2018prl, Hebatalla2019aml}. 
	Recent improvement of the energy resolution reveals fine features due to small distortions~\cite{vanSchooneveld2013acie, Liu2016ic, Huang2017nc, Wang2019prb} and the spin-orbit coupling~\cite{vanSchooneveld2012jpcc}. 
	In combination with the theoretical simulations, Co $2p3d$ RIXS provides accurate details about the electronic structures of both the ground state and the excited states with chemical site selectivity.
	
	\begin{figure}[b]
		\centering
		\includegraphics[width=0.95\columnwidth]{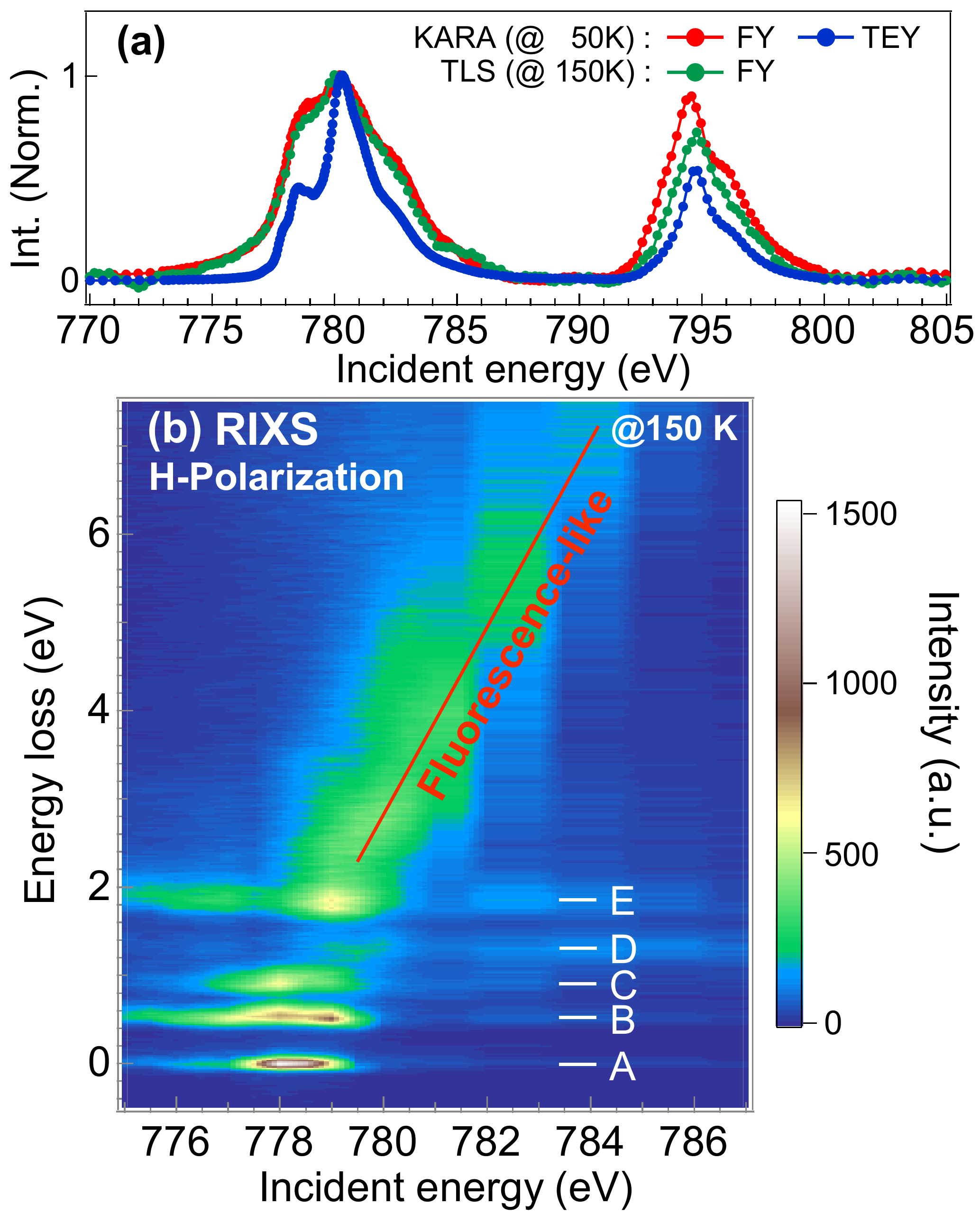}
		\caption{The experimental results. (a) The $H$-polarization $2p$ XAS spectra of Co$_{3}$O$_{4}$. (b) The $H$-polarization $2p3d$ RIXS energy map of Co$_{3}$O$_{4}$.}\label{exp_XAS_RIXS}
	\end{figure}
	
	\section{Methodology}
	\subsection{Experimental details}
	The experiments were performed on 99.9985\% pure Co$_3$O$_4$ powder produced by Alfa Aesar. 
	A cylindrical pellet Co$_3$O$_4$ with 10 mm in diameter and 0.5 mm in height was prepared for the measurements. 
	Co $2p$ XAS spectra were acquired at the soft X-ray WERA beamline of the Karlsruhe Research Accelerator (KARA) synchrotron in Germany. 
	The instrumental resolution was calibrated to be $\sim$ 280 meV full width at half-maximum (FWHM) at the Co $L_{2,3}$ edge ($\sim$780 eV). 
	Both the total electron yield (TEY) and the fluorescence yield (FY) methods were employed.
	The Co $2p3d$ RIXS measurements with linearly vertical ($V$) and horizontal ($H$) polarized incident X-rays were performed at the Taiwan Light Source (TLS) beamline 05A of National Synchrotron Radiation Research Center (NSRRC) in Taiwan~\cite{Lai2014jsr}. 
	The experimental energy resolution of the incident photon was $\sim$700 meV and the combined resolution of RIXS was $\sim$ 90 meV. 
	Here we note that the RIXS resolution is much better than the incident photon resolution thanks to the position sensitive detector combined with the monochomator-spectrometer system based on the energy-compensation principle~\cite{Lai2014jsr, Fung2004aip}.
	A grazing incident geometry ($\sim20^\circ$) with the spectrometer at 90$^\circ$ was used. 
	To compare the XAS spectra measured with different facilities and approaches, the background signals were subtracted from the spectra, as described in the Supplementary Material (SM).  
	The subtracted spectra were normalized at the peak with the maximum intensity at the Co $L_3$ edge. 
	We calibrated the photon energy of the RIXS beamline to the spectra acquired at the WERA beamline.  
	The RIXS spectra measured with the $H$-polarization were normalized for the exposure time. Then, the spectra with $V$-polarization were normalized to $H$-polarization ones according to the profile at high energies (above 2.5~eV).
	The measurements in the KARA-WERA and TLS-05A beamlines were carried out at 50 K and 150K, respectively.  
	
	\subsection{Simulations}
	We analyze the experimental data using the cluster model which includes the Coulomb multiplet interaction, the crystal-field splitting, and the spin-orbit coupling on the X-ray excited Co $3d$ site as well as a charge transfer between Co $3d$ and O $2p$ orbitals on the adjacent sites. 
	The many-body Hamiltonian of the cluster model is solved using the Quanty program, which implements the configuration-interaction scheme~\cite{deGroot2005ccr, Haverkort2012prb}. 
	To study the mixed-valence state of Co$_3$O$_4$, we use two cluster models simulating the spectra of the Co$^{2+}$(T$_d$) site and the Co$^{3+}$(O$_h$) site. 
	We take the initial parameter values of the cluster models from previous studies for cobaltate with a high-spin ground state ($^4$B$_{1}$ in D$_{2d}$ symmetry) and a low-spin ground state ($^{1}$A$_{1g}$ in O$_{h}$ symmetry)~\cite{Liu2016ic, Tomiyasu2017prl, Wang2019prb}.
	The parameter values are refined to represent the studied compound by a detailed comparison with the present high-resolution RIXS data and its polarization dependence, which will be discussed in Sec.~\ref{sim_result}. 
	To evaluate the used parameter values, we present the ones estimated by an $ab$-initio calculation for the real crystal structure of Co$_3$O$_4$. Starting with a density-functional calculation with local-density approximation (LDA), we construct a tight-binding model spanning Co $3d$ and O $2p$ bands by a Wannier projection. The parameter values (crystal-field and hopping parameters) are extracted in the tight-binding model. 
	
	\section{Results}
	\subsection{Experimental results}
	Figure~\ref{exp_XAS_RIXS} shows the experimental Co $2p$ XAS and $2p3d$ RIXS results. 
	The TEY spectrum shows a sharp feature at 778.5~eV and at  780.2~eV which are characteristic of the absorption by the Co$^{2+}$ and the Co$^{3+}$ sites in Co$_3$O$_4$, respectively~\cite{Hibberd2015jpcc}. 
	The FY spectrum, on the other hand, is rather broad and the features are obscure.
	This is due to strong saturation and self-absorption effects for a bulk sample, which are more effective for features with strong absorption intensities. 
	In the $2p3d$ RIXS result (Fig.~\ref{exp_XAS_RIXS}b), a fluorescence-like signal is observed and its excitation energy increases with incident photon energies as guided by a line.
	In addition, sharp features at $\sim$0.0, 0.5, 0.9, 1.2, and 1.9 eV (labelled by A to E in Fig~\ref{exp_XAS_RIXS}b) are observed. 
	The features A--C are resonantly enhanced at 778 eV incident photon energy, while the feature D is enhanced at $\sim$780 eV. The feature E is resonated at the energy slightly below 779 eV. 
	Those features are attributed to local excitations of the Co$^{2+}$ and/or Co$^{3+}$ sites, as we will discuss later.
	
	\begin{figure}[t]
		\centering
		\includegraphics[width=0.9\columnwidth]{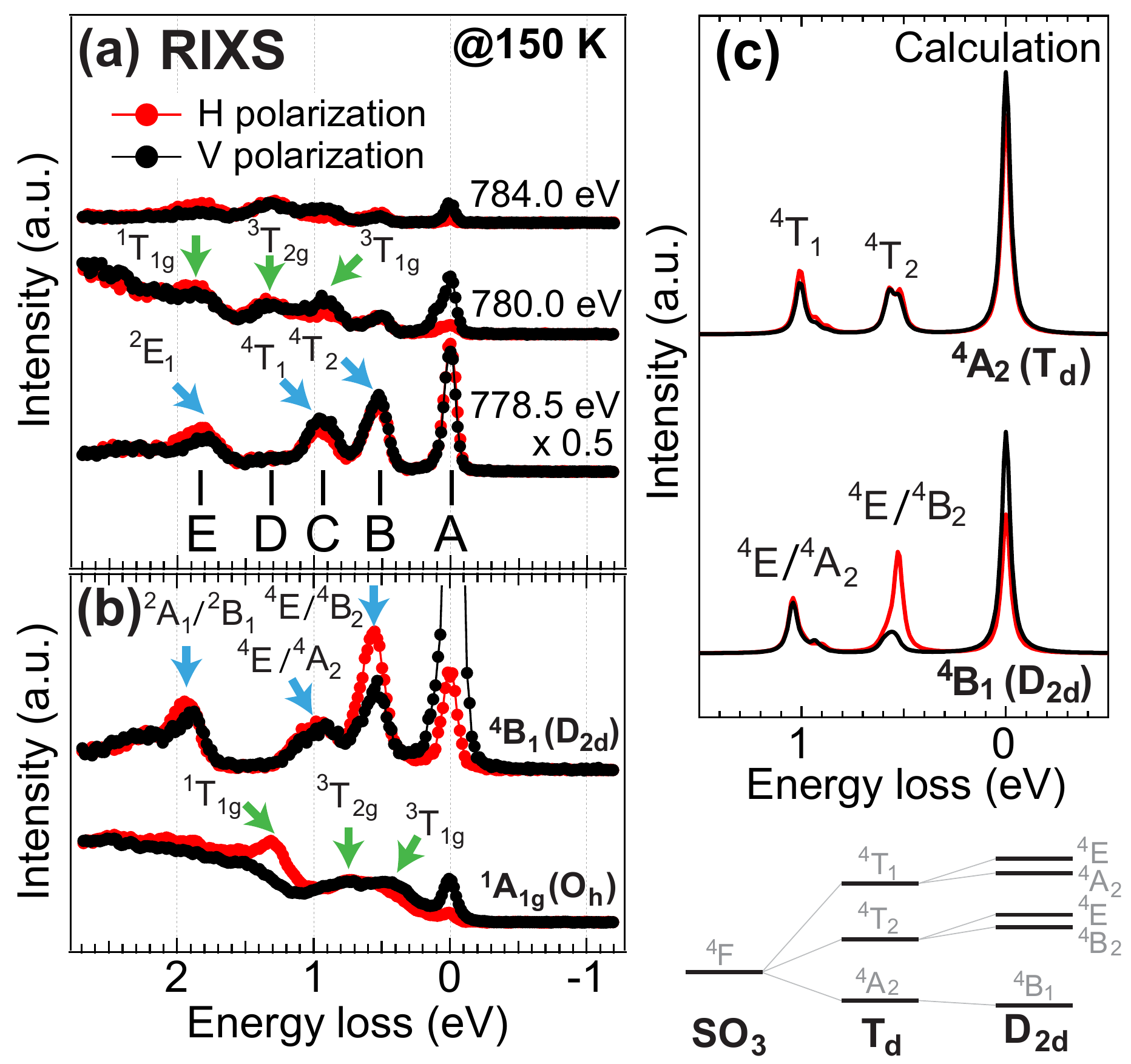}
		\caption{The comparison of $H$- and $V$-polarization RIXS spectra of (a) the Co$_{3}$O$_{4}$ and (b) the references~\cite{Liu2016ic, Wang2019prb}. 
			The references are the spectra of the $^{4}$B$_{1}$(D$_{2d}$) and $^{1}$A$_{1g}$(O$_{h}$) ground states, which were taken on the K$_{5}$H[CoW$_{12}$O$_{40}$]$\cdot$xH$_{2}$O at 300 K~\cite{Liu2016ic} and on the LaCoO$_3$ crystal at 20 K~\cite{Wang2019prb}, respectively. 
			The blue (green) arrows indicate the characteristic features of the  Co$^{2+}$ (Co$^{3+}$) site.
			(c) The calculated $2p3d$ RIXS polarization comparison of distorted and non-distorted tetrahedral Co$^{2+}$. The states splitting caused by the symmetry broken from a SO$_3$ symmetry through the T$_{d}$ symmetry to the D$_{2d}$ symmetry is presented below. The calculation was carried out by applying the parameters in reference~\cite{Liu2016ic}, where ligand-to-model charge transfer channel was not included.}\label{exp_XAS_RIXSexp_poldep}
	\end{figure}
	
	To gain more insight, figure~\ref{exp_XAS_RIXSexp_poldep}a compares the $H$- and $V$-polarization spectra. 
	At 780.0~eV, the feature C (at $\sim$0.9 eV) is enhanced by the $V$-polarization, while the feature E (at $\sim$1.9 eV) shows opposite behaviour. 
	This polarization dependence resembles the one of $^3$T$_{1g}$ and $^1$T$_{1g}$ excited states in LaCoO$_3$ at 20 K, where the Co$^{3+}$ ions have the $^1$A$_{1g}$ ground state, see Fig.~\ref{exp_XAS_RIXSexp_poldep}a and ~\ref{exp_XAS_RIXSexp_poldep}b. 
	However, the excitation energies differ largely in Co$_3$O$_4$ and LaCoO$_3$, indicating that the crystal field splitting varies substantially between the two. 
	In contrast, the feature B (at $\sim$0.5 eV) is identified as the $^4$T$_2$ excited state on the Co$^{2+}$(T$_d$) site~\cite{Mironova1994ssc}.
	As a reference, Fig.~\ref{exp_XAS_RIXSexp_poldep}b shows Co $L_3$ RIXS in K$_5$H[CoW$_{12}$O$_{40}$]$\cdot$xH$_{2}$O~\cite{Liu2016ic} with a divalent Co$^{2+}$ ion and $^4$B$_{1}$ (D$_{2d}$) ground state in a distorted tetrahedral structure. 
	In this reference compound, the distortion changes the ground state symmetry from the $^4$A$_2$(T$_d$) to the $^4$B$_{1}$(D$_{2d}$) that gives rise to a strong polarization dependence at 0.5 eV peak in the RIXS spectra.
	However, the feature B shows no dichroism in Co$_3$O$_4$, which suggests that the distortion is negligibly small on the Co$^{2+}$ site and supports the $^4$A$_2$(T$_d$) symmetry of the ground state in the calculation.
	
	\begin{figure}[t]
		\centering
		\includegraphics[width=0.85\columnwidth]{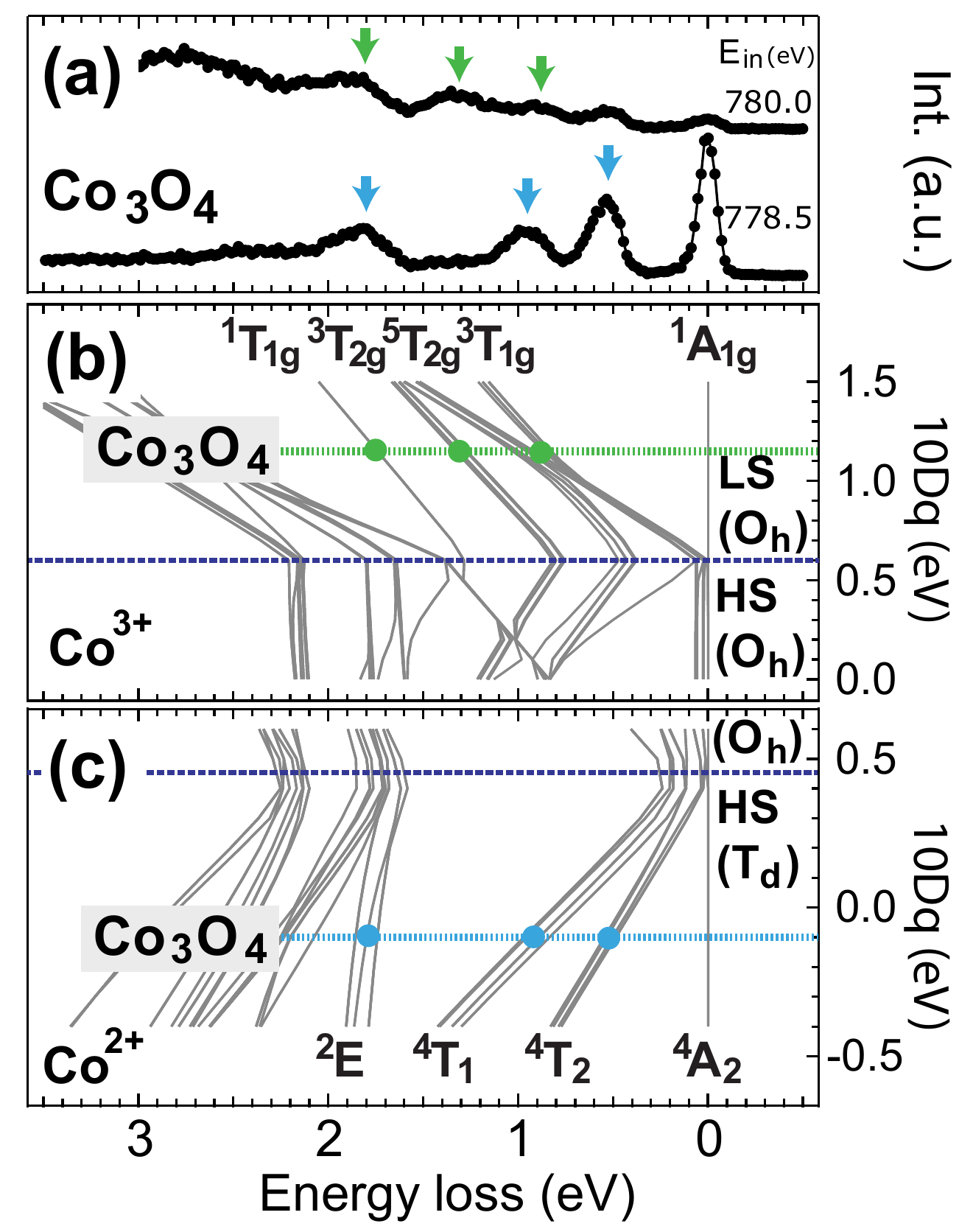}
		\caption{(a) The $H$-polarization $2p3d$ RIXS spectra of Co$_{3}$O$_{4}$ excited at 778.5 and 780.0 eV. The calculated energy diagrams of (b) the Co$^{3+}$ ion and (c) the Co$^{2+}$ ion as a function of crystal field energy ($10Dq$) including the charge transfer.
		}\label{parameter}
	\end{figure} 
	
	\subsection{Simulated results}\label{sim_result}
	Guided by the observed low-energy features, we determine the low-energy local excitations on the two Co sites in Co$_3$O$_4$. 
	In the cluster-model analysis, the crystal-field parameter ($10Dq$) represents the energy splitting of Co 3$d$ orbitals ($t_{2g}$($t_{2}$) and $e_g$($e$) orbitals in O$_h$(T$_d$) symmetry). 
	Figures~\ref{parameter}b and \ref{parameter}c show the excited states energies as a function of the $10Dq$ for the Co$^{3+}$ and Co$^{2+}$ site, respectively. 
	The $2p3d$ RIXS probed at 780.0~eV shows three features at 0.9, 1.3, and 1.9 eV, indicated with the green arrows in Fig.~\ref{parameter}a, which are attributed to the Co$^{3+}$-site signals. 
	As seen in Fig.~\ref{parameter}b, the matching of the three features gives $10Dq$ $\sim$1.15 eV at the Co$^{3+}$ site.  
	The $10Dq$ value is far larger than the spin-state transition point ($10Dq\sim$~0.6~eV), suggesting that the ground state on the Co$^{3+}$ site in Co$_3$O$_4$ is a robust low-spin singlet $^1$A$_{1g}$. 
	For the Co$^{2+}$ site, the $10Dq$ $\sim-0.1$~eV reproduces the energy position of the characteristic features at 0.5, 0.9, 1.9~eV in the RIXS data probed at 778.5~eV, compare Figs.~\ref{parameter}a and \ref{parameter}c.
	The negative $10Dq$ value indicates an inversion of the $e$ and $t_{2}$ manifolds in the T$_d$ symmetry. 
	
	\begin{table}[b]
		\caption{The model parameters used in the simulation (in eV), which are the crystal field energy, hopping integrals, charge transfer energy, U$_{dd}$ and U$_{pd}$. The \textit{i} and \textit{m} stand for the configurations of initial ground state and intermediate state, respectively. }\label{tab_Co3O4_parameter}
		\setlength{\extrarowheight}{3pt}
		\centering
		\begin{tabular}{l c c c c c c c }
			\hline
			\hline
			& $10Dq$ & $10Dq_{\rm eff}$ & $\Delta$ & V$_{e({e_g})}$ & V$_{t_{2}(t_{2g})}$ & U$_{dd}$ & U$_{pd}$ \\
			\hline
			\hspace{5pt}Co$^{2+}_{i}$ & $-$0.10 & $-$0.55 & 4.5 & 1.0 &  2.0  & 4.5  &  - \\
			\hspace{5pt}Co$^{2+}_{m}$ & $-$0.02 & $-$0.47 & 4.5 & 1.0 &  2.0  & 4.5  & 6.0 \\
			\hline
			\hspace{5pt}Co$^{3+}_{i}$ & 1.15 & 1.90 & 1.5 & 3.12 & 1.8  & 6.5  &  - \\
			\hspace{5pt}Co$^{3+}_{m}$ & 0.84 & 1.59 & 1.5 & 3.12 & 1.8  & 6.5  & 7.5 \\
			\hline
			\hline
		\end{tabular}
	\end{table}
	
	Table~\ref{tab_Co3O4_parameter} summarizes the used parameter values. 
	The charge transfer energy $\Delta$ is an energy related to electron transfer from a ligand to the Co site and  V$_{e({e_g})}$/V$_{t_{2}(t_{2g})}$ are the values for electron hopping. 
	The U$_{dd}$ and U$_{pd}$ values parameterize the Coulomb interaction, which are set to reference values~\cite{Tomiyasu2017prl, Wang2017jpcc}. 
	Since the ligand-to-metal charge transfer also contributes to the energy splitting of Co 3$d$ states, we provide the $10Dq_{\rm eff}$ value which is calculated by the resultant energy splitting by including the charge transfer.  
	To simulate a contraction of Co $3d$ wave functions by the presence of the core hole~\cite{Cramer1991jacs}, the $10Dq_{\rm eff}$ value in the intermediate state is reduced from that in the ground state by $\sim$15\%, see SM. 
	
	\begin{figure}[t]
		\centering
		\includegraphics[width=0.95\columnwidth]{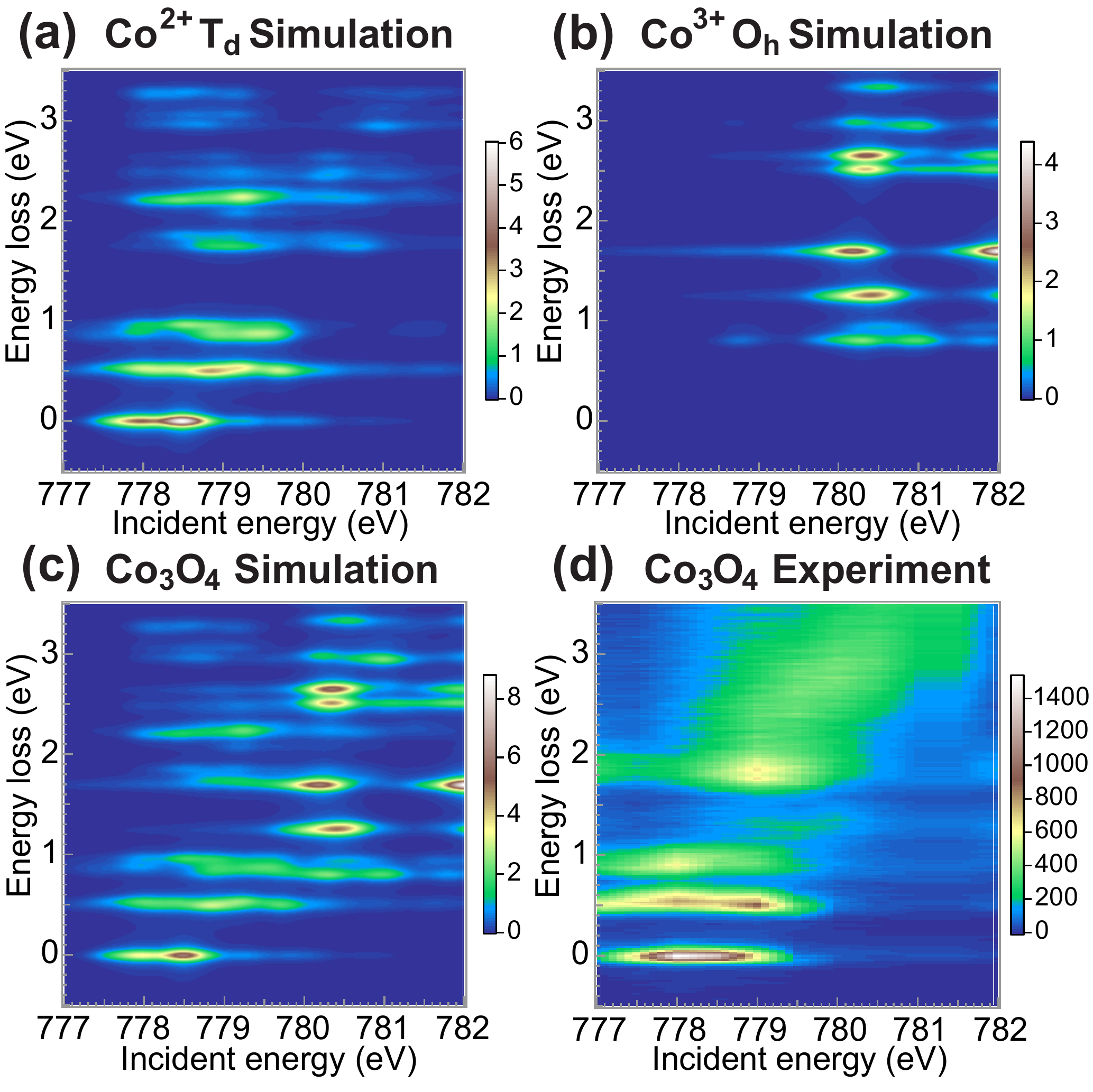}
		\caption{The simulated results of $2p3d$ RIXS energy maps. (a) Co$^{2+}$ $^{4}$A$_{2}$ (T$_{d}$) ground state and (b) Co$^{3+}$ $^{1}$A$_{1g}$ (O$_{h}$) ground state. 	
			(c) The total $2p3d$ RIXS simulated results obtained by the summation of two different sites using a weighting 3:8. (d) The result of Co$_3$O$_4$. Here the $H$-polarization is applied in all the results.
		}\label{exp_RIXS_calvsexp}
	\end{figure}
	
	To evaluate the values in Table~\ref{tab_Co3O4_parameter}, the parameter values estimated by the LDA calculation are provided. 
	For the Co$^{2+}$($T_d$) site, the estimated values are $-$0.10~eV for $10Dq$ and 1.29~(1.82) eV for $V_{e}$ ($V_{t_{2}}$).
	For the Co$^{3+}$(O$_h$) site, the estimated values are 0.7~eV for $10Dq$ and 3.03~(1.74) eV for $V_{e_{g}}$ ($V_{t_{2g}}$)~\footnote{At the Co$^{3+}$ site, the $t_{2g}$ manifolds split into a singlet and doublet due to a small trigonal distortion. The $V_{t_{2g}}$ value is estimated by averaging the hopping integrals over the two.}. 
	The $D\sigma$ value, which measures the trigonal distortion, is estimated as 0.05~eV. 
	This value is much smaller than the required value ($>$ 0.5 eV) for changing the ground state symmetry (singlet $^1$A$_{1g}$) of the Co$^{3+}$ site in Co$_3$O$_4$, but gives a minor correction to multiplet energies. Since the change affects minor on the spectra, we thus neglect it in our simulation and the site is referred to as O$_h$ for simplicity. 
	Overall the optimized values in the cluster-model simulation agree well with the $ab$-initio estimates. 
	The small discrepancy in the $10Dq$ value at the Co$^{3+}$ site is likely due to an underestimate of the covalency in the LDA scheme\cite{Haverkort2012prb,Hariki2020prb}.
	
	Figures~\ref{exp_RIXS_calvsexp}a and \ref{exp_RIXS_calvsexp}b show the RIXS intensities calculated for the Co$^{2+}$ and Co$^{3+}$ sites, respectively, using the optimized parameters in Table~\ref{tab_Co3O4_parameter}.
	Since the low-spin ($^1$A$_{1g}$) ground state on the Co$^{3+}$ site is angular isotropic, its intensity is strongly suppressed when the scattering angle is about 90$^\circ$ with the $H$-polarization condition. 
	Thus the intensity at zero energy loss (elastic line) is mainly due to the Co$^{2+}$ site with the $^4$A$_{2}$ ground state. 
	Figure \ref{exp_RIXS_calvsexp}c shows the sum of the intensities of the two sites weighted by the ratio of 3:8 for the normalized (to total area) Co$^{2+}$ and Co$^{3+}$ spectra.
	The ratio simulates the combination of stoichiometry (1:2) and the number of holes (3:4). The incident energies of Co$^{2+}$ and Co$^{3+}$ spectra were adjusted by a shift of 1.8 eV to reproduce the experimental Co $L_3$ XAS (cf Fig.~\ref{exp_PFY}a). 
	The cluster-model simulation above identifies the site-resolved excitation spectra in Co$_3$O$_4$. To reproduce the RIXS intensities, the saturation and the self-absorption effects need to be taken into account.
	
	\section{Discussion}
	\subsection{Site selectivity of partial fluorescence yield spectra}
\begin{figure}[t]
		\centering
		\includegraphics[width=0.90\columnwidth]{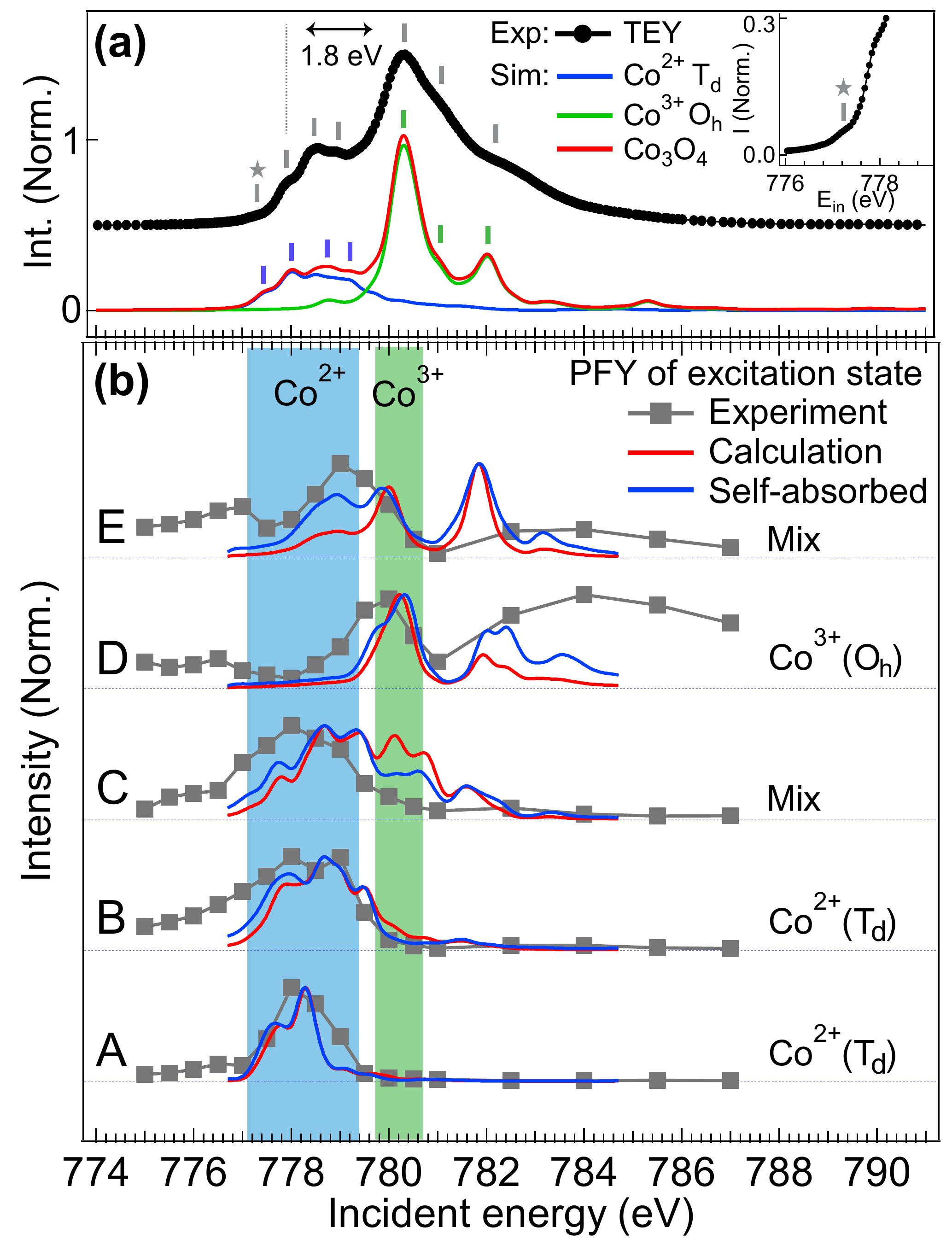}
		\caption{The comparison of XAS spectra. (a) The comparison between the experimental TEY spectra of the Co$_3$O$_4$ and the simulation results using the optimized parameters. The inset panel zooms in the pre-edge region and indicates a small feature before the edge. This small feature matches to the first feature indicated in the simulation. (b) The comparison of the PFY spectra correspond to the excited states A--E.
		}\label{exp_PFY}
	\end{figure}

	We extracted the partial fluorescence yield (PFY) spectra to see the local excitations in detail. 
	Before the discussion on the PFY spectra, the simulated XAS result is presented in comparing with the experimental data. 
	In Fig.~\ref{exp_PFY}a, the sum of the simulated XAS of Co$^{2+}$ (blue) and Co$^{3+}$ (green) sites well matches with the experimental TEY data (gray bars and red line). 
	The contributions of two sites are separated in the incident energies, that enables to observe site-resolved local excitations by RIXS, as we shown in Fig.~\ref{exp_RIXS_calvsexp}. 
	Details of the local excitations can be stressed in PFY spectra, which are acquired for energy transfer of the $dd$ excitations (A--E).
	The experimental data are obtained by fitting the RIXS intensities of these $dd$ excitations (cf Fig.~\ref{exp_RIXS_calvsexp}d), and the theoretical spectra are obtained by summing simulated intensities in a narrow energy window centered at these excitations in Fig.~\ref{exp_RIXS_calvsexp}c.
	In the PFY spectra (Fig.~\ref{exp_PFY}b), the features A--B and D are unambiguously attributed to the Co$^{2+}$-site and the Co$^{3+}$-site excitations, respectively, while the features C and E show an overlap with both Co$^{2+}$ and Co$^{3+}$ regions. 
	We stress that the overlap is essential since, according to our simulation in Fig.~\ref{exp_RIXS_calvsexp}a and \ref{exp_RIXS_calvsexp}b, both two sites have excitations at around 0.9 and 1.9~eV. 
	To discuss the two features, however, saturation and self-absorption effects needs to be taken into account properly.
	
	A systematic discrepancy is observed that the intensity at about 780 eV were always overestimated in the simulation, particularly, of the PFY spectra of the features C and E. 
	This indicates that the spectral intensity with respect to the Co$^{2+}$ site is more pronounced. 
	To confirm the statement, we applied the saturation and self-absorption corrections as~\cite{Achkar2011prb, Wang2020jsr}:
	\begin{equation}
	\begin{aligned}
	I_{ems} = I_0 \frac{S_X(\omega_{in}, \omega_{out})\mu_X(\omega_{in})}{\mu(\omega_{in}) + \mu(\omega_{out})\frac{sin\theta}{sin(\alpha - \theta)}}.
	\label{eq_selfabs}
	\end{aligned}
	\end{equation}
	Here, $X$, $\alpha$, and $\theta$ refer to the emission edge of element, the scattering angle, and the sample rotation angle, respectively. 
	In this work, we employed the experimental geometry with $\alpha=90^\circ$ and $\theta=20^\circ$. 
	The $\mu(\omega_{\rm in}$) and $\mu(\omega_{\rm out}$) are the absorption factor of \textit{photon-in} and \textit{photon-out} channel in the RIXS process, see further details about the background in the SM and the reference~\cite{Wang2020jsr}. 
	We multiply a self-absorption coefficient ($\mu(\omega_{\rm in}) + \mu(\omega_{\rm out})\frac{{\rm sin}\theta}{{\rm sin}(\alpha - \theta)}$)$^{-1}$ to the simulated RIXS result (multiplication of $S_X(\omega_{\rm in}, \omega_{\rm out}$) and $\mu_X(\omega_{\rm in}$)). 
	The formula implies that the saturation effect is more strong for a large absorption factor. 
	Thus, the features at about 780 eV in PFY spectra are to be suppressed. 
	Consequently, the PFY weights on the Co$^{3+}$ region is largely suppressed for the feature C (cf. Fig.~\ref{exp_PFY}b).  
	It also enhances the intensity at $\sim$782 eV of PFY spectra for the feature D and shows better agreement with the experiment. 
	In the simulations, the energy broadening was assumed to be the same for both sites, which yields rather sharp features in the Co$^{3+}$-site contribution compared to the experiment.
	A larger energy broadening for the Co$^{3+}$ site is caused by the strong ligand-metal hybridization on the Co$^{3+}$ site.
	To reproduce the broad XAS/PFY structure above the $L_3$~edge, the band formation of ligand 2$p$ states must be taken into account, which is beyond the description of the cluster model used in this study.
	
	The present analysis shown that the excited states at 0.5 eV and 1.3 eV are the unique features to identify the $^4$T$_{2}$(T$_d$) excited state on the Co$^{2+}$ site and $^3$T$_{1g}$(O$_h$) excited state on the Co$^{3+}$ site. 
	Furthermore, to the best of our knowledge, the $^4$T$_2$ excited state of the Co$^{2+}$ site in Co$_3$O$_4$ was only reported by Mironova \textit{et al.} through infrared more than two decades ago~\cite{Mironova1994ssc}. 
	The result confirms that the Co$_3$O$_4$ is mainly composed by the magnetically active high-spin Co$^{2+}$(T$_d$) and the diamagnetic low-spin Co$^{3+}$(O$_h$). 
	
	\begin{figure}[t]
		\centering
		\includegraphics[width=0.98\columnwidth]{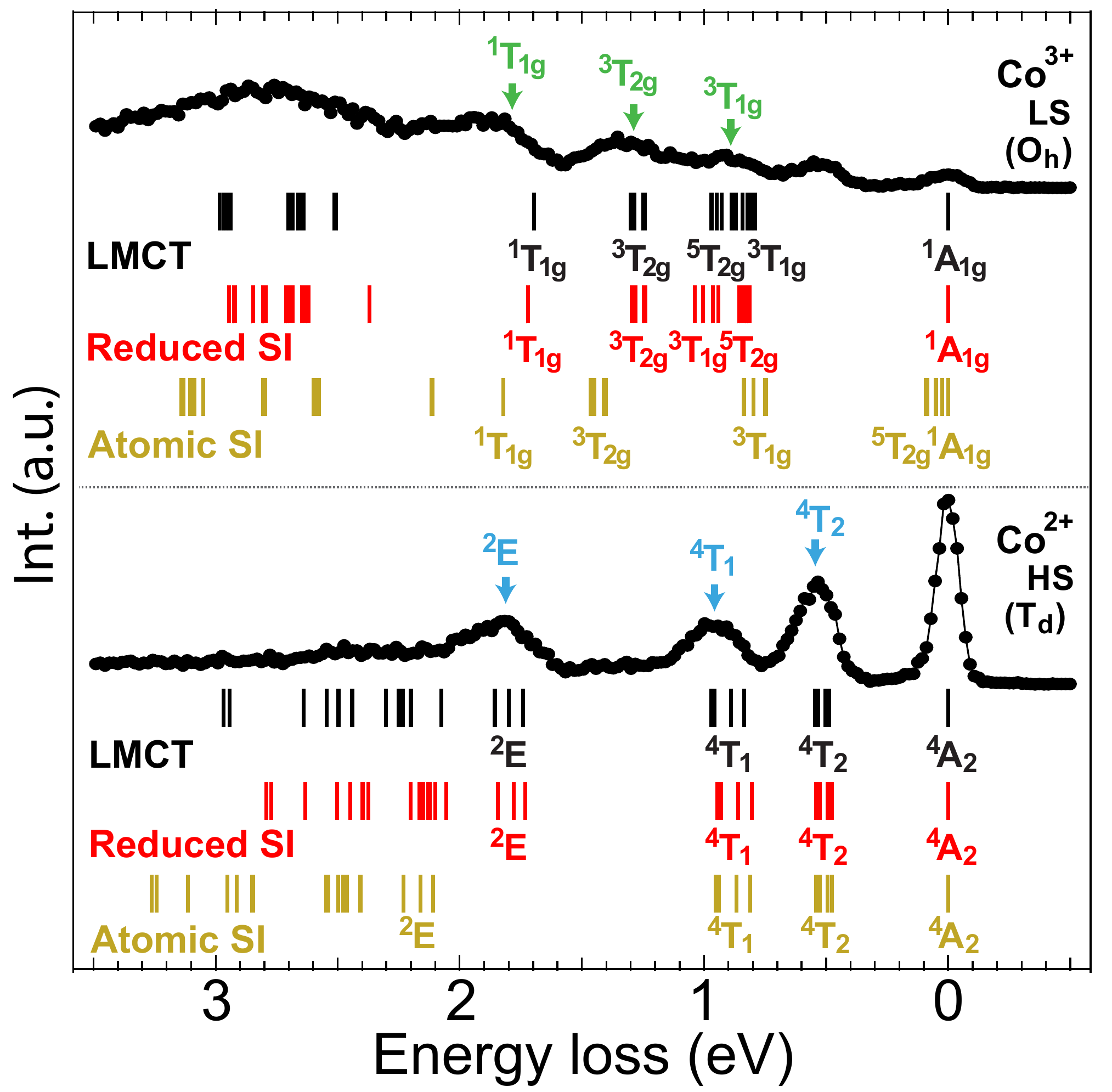}
		\caption{The energy diagrams calculated using the values of the atomic Slater integrals (atomic SI), the reduced Slater integrals (reduced SI), and the ligand-to-metal charge transfer (LMCT) effect. 
		}\label{state_comparing}
	\end{figure}
	
	\subsection{Ligand-metal hybridization influence of different Co sites in Co$_3$O$_4$}
	We discuss the question whether the ligand-metal hybridization influences the local electronic structure based on the ligand-to-metal charge transfer model. 
	Hibberd~\textit{et al.}~have shown in their atomic-model analysis that the Slater integrals of the atomic Coulomb multiplet need to be reduced substantially to reproduce the Co $L_3$ XAS spectra, which implies that the ligand-metal hybridization is strong in Co$_3$O$_4$~\cite{Hibberd2015jpcc,Chen2011prb, Qiao2013jmcc}.
	The sensitivity of $2p3d$ RIXS to the $dd$ excitations allows us to address the question, and furthermore study the site-dependence of the covalency in Co$_3$O$_4$. 
	Figure~\ref{state_comparing} compares the energy diagrams obtained with three different models: (i) the atomic model with the bare (atomic) values of the Slater integrals, (ii) the one with reduced Slater integrals, and (iii) the cluster model including the ligand-to-metal charge transfer channel explicitly. 
	Apparently, the atomic model (i) overestimates energies of observed excitations.
	For the Co$^{2+}$ site, the atomic model (ii) with reduced Slater integrals (80\% from the atomic values) yields good agreement with the experimental data. 
	On the other hand, for the Co$^{3+}$ (O$_h$) site, the Slater integrals are reduced to 55\% (80\%) from the atomic F${^\mathrm{2}_{dd}}$ (F${^\mathrm{4}_{dd}}$) value to fit the experimental data. 
	The unconventional reduction rates for the Co$^{3+}$ (O$_h$) site imply that the screening effect via a ligand-to-metal charge transfer channel is not negligible. 
	The cluster model including the ligand-to-metal charge transfer channel shows good agreement to the experimental results.
	
	To obtain further information about the ligand-metal hybridization, Table~\ref{tab_Co3O4_weight} shows the configuration weights in the ground state of the cluster model using the optimized parameters in Table~\ref{tab_Co3O4_parameter}. 
	The ligand-to-metal charge transfer channel mixes the ionic configuration (3$d^n$) with ones with ligand holes (3$d^{n+1}\underline{L}^1$ and $3d^{n+2}\underline{L}^2$). 
	In the Co$^{3+}$ site, the ligand-hole configurations (3$d^{7}\underline{L}^1$ and $3d^8\underline{L}^2$) show large weights, which indicates that the Co$^{3+}$ site is strongly hybridized with the ligand states. 
	In contrast, the ligand-hole configuration (3$d^{8}\underline{L}^1$) contributes only $\sim$20\% to the ground state in the Co$^{2+}$ site. 
	This observation suggests that the Co$^{3+}$ site is highly covalent, while the Co$^{2+}$ site is rather ionic in Co$_3$O$_4$. 
	The difference also affects the RIXS profile: the ionic Co$^{2+}$ site shows sharp local $dd$ excitations; the covalent Co$^{3+}$ site exhibits a broad intense fluorescence-like feature (cf. Fig.~\ref{exp_XAS_RIXS}).
	
	The orbital covalency of the Co$^{2+}$ (Co$^{3+}$) cation can be analyzed using the approach described in the SM and literature~\cite{Wang2017jpcc}. 
	The orbital covalencies of e and t$_{2}$ orbitals on the tetrahedral Co$^{2+}$ cation in Co$_3$O$_4$ are 100\% and 80\%.
	The e orbital is fully occupied and cannot participate in the ligand-metal hybridization, thus 100\% orbital covalency is found. 
	A high value of the cation orbital covalency for t$_2$ orbital indicates that it less contributes to the ligand-hole configuration $d^8\underline{L}$, which is consistent with the ionic character of the Co$^{2+}$ site.
	For the Co$^{3+}$ site, the orbital covalencies of e$_g$ and t$_{2g}$ orbitals are 50\% and 100\%.  
	This indicates that the ligand-metal hybridization mainly influence to the e$_g$ orbital of the Co$^{3+}$ ions. 
	Although the Co$^{3+}$ cation is the singlet ground state ($t^6_{2g}e_g^0$ state in the atomic picture), the $e_g$ orbital forms a strong bonding with neighboring oxygen 2$p$ orbitals, which involves the $d^7\underline{L}$ configuration with a ligand hole with e$_g$ symmetry, yielding the reduced value (50\%) of the orbital covalency. 
	
	\begin{table}[t]
		\caption{The weight of configurations and cation orbital covalency in ground state (unit in \%). Although the number of ligand holes is considered up to two in the spectral simulations, the covalency is estimated only using the configurations up to one ligand hole.}\label{tab_Co3O4_weight}
		\centering
		\begin{tabular}{l c c c c c}
			\hline
			\hline
			& \multicolumn{3}{c}{weight of configurations} & \multicolumn{2}{c}{orbital covalency}\\
			& $\ket{3d^n}$ & $\ket{3d^{n+1}\underline{L}^1}$ & $\ket{3d^{n+2}\underline{L}^2}$ & e(e$_{g}$) & t$_{2}$(t$_{2g}$) \\
			\hline
			\hspace{5pt}Co$^{2+}$($3d^7$) & 79  & 20 &   1  & 100  & 80 \\
			\hspace{5pt}Co$^{3+}$($3d^6$) &  40  & 50 & 10  & 50  & 100 \\
			\hline
			\hline
		\end{tabular}
	\end{table}

	\section{Conclusion}
	We present the Co $2p$ XAS and $2p3d$ RIXS experimental results in comparison with cluster model simulations. 
	The $2p3d$ RIXS provides good chemical site selectivity to the local electronic structure, from which we are able to identify orbital covalencies of different ions in the compound. 
	The polarization dependent analysis indicates the symmetry character of the $dd$ excitations, which provides a solid guide to analyze the local electronic structure. 
	By selecting characteristic excitations for the $2p3d$ RIXS spectra, the PFY spectra have the ability to provide additional site dependent information. 
	The result shows the $^4$T$_{2}$ excited state of the tetrahedral Co$^{2+}$ site at 0.5 eV, which is beyond the discriminative power of optical absorption. 
	In addition, the $^1$A$_{1g}$ to $^3$T$_{2g}$ excitation of the octahedral Co$^{3+}$ site at 1.3 eV can be uniquely identified. 
	The ground state electronic structure of the Co$^{2+}$ ions and the Co$^{3+}$ ions are respectively high-spin $^4$A$_{2}$(T$_d$) and low-spin $^1$A$_{1g}$(O$_h$), where the high-spin Co$^{2+}$ must be the magnetically active site.
	Our result also shows strong ligand-metal hybridization on the Co$^{3+}$ site, 
	which indicates that the Co$^{3+}$ site in Co$_3$O$_4$ is rather covalent. 
	In contrast, the Co$^{2+}$ site shows weak hybridization implying that Co$^{2+}$ is more ionic. 
	This chemical site selectivity will help the further understanding on the site-dependent catalytic activity and magnetic activity of the spinel cobalt oxides.
	
	\section{Acknowledgements}
	We gratefully acknowledge the synchrotron light source KARA and the KNMF at Karlsruhe, Germany, and the Taiwan Light Source at Hsinchu, Taiwan, for the provision of beamtime. 
	The authors thank the technical staff for their help with the XAS and RIXS measurements. The experiments were supported by ERC advanced grant (grant agreement No. 340279-XRAYonACTIVE). 
	D.J.H. was supported by the Ministry of Science and Technology of Taiwan under Grant No. 106-2112-M-213-008-MY3. 
	
	\bibliography{ms}
	
\clearpage
	
\end{document}


\title{Supplementary Material of ``Revealing the site selective local electronic structure of Co$_{3}$O$_{4}$ using $2p3d$ resonant inelastic X-ray scattering"}
	
	\author{Ru-Pan Wang}
	\author{Meng-Jie Huang}
	\author{Atsushi Hariki}
	\author{Jun Okamoto}
	\author{Hsiao-Yu Huang}
	\author{Amol Singh}
	\author{Di-Jing Huang}
	\author{Peter Nagel}
	\author{Stefan Schuppler}
	\author{Ties Haarman}
	\author{Boyang Liu}
	\author{Frank M. F. de Groot}
	
	
	\maketitle
	\renewcommand{\theequation}{S.\arabic{equation}}
	\renewcommand{\thefigure}{S.\arabic{figure}}
	\renewcommand{\thetable}{S.\arabic{table}}
	\renewcommand{\bibnumfmt}[1]{[s#1]}
	\renewcommand{\citenumfont}[1]{s#1}
	
	\subsection{$2p$ XAS spectra background subtraction}
\begin{figure}[h]
		\centering
		\includegraphics[width=0.85\columnwidth]{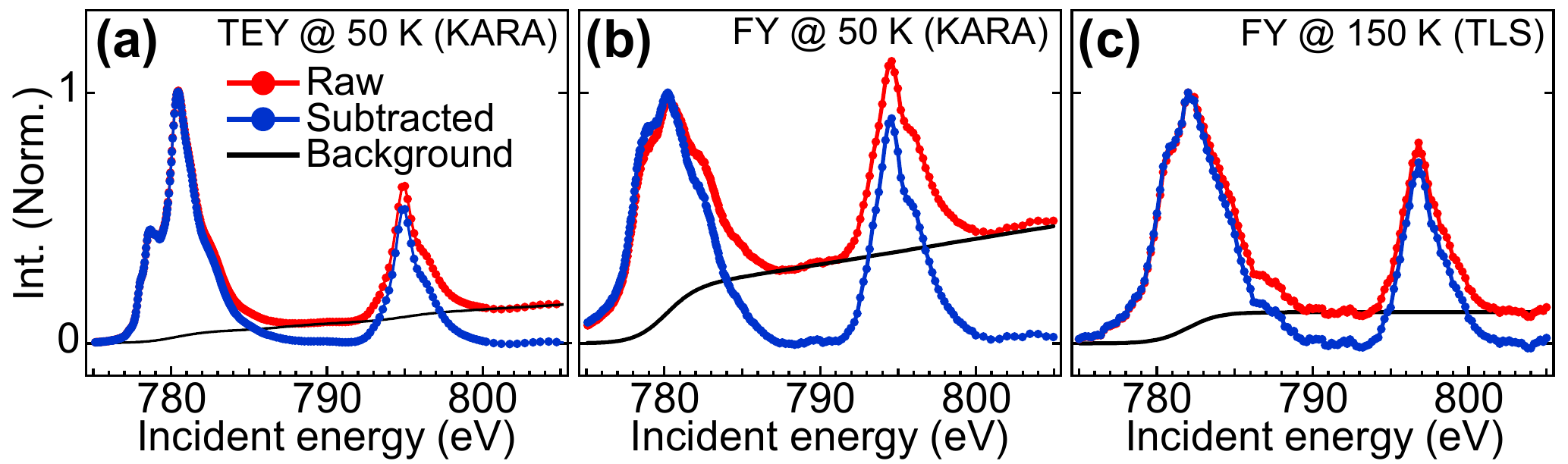}
		\caption{The data treatment of XAS spectra. (a) The TEY-XAS and (b) the FY-XAS spectra measured at WERA beamline in KARA. (c) The FY-XAS spectra measured at soft X-ray RIXS beamline in TLS.}\label{exp_XAS_treatment}
	\end{figure}

	Figure~\ref{exp_XAS_treatment} shows the raw $2p$ XAS spectra (red), subtracted $2p$ XAS spectra (blue), and the background profile (black). 
	We subtracted the background signal from the original XAS results, where the background signal contains edge jump(s), particles scattering, and linear signal. 
	The subtracted spectra were normalized to the maximum of the Co $L_3$-edge. 
	The photon energy of RIXS beamline were calibrated to the spectra acquired in WERA beamline, where the calibration also applied to the incident energy of RIXS spectra.
	
	\subsection{Theoretical absorption background estimation}
	According to the tabulated data~\cite{Henke1993DataBase}, we can estimate the attenuation length for the individual elements of Co$_3$O$_4$ ($\rho$= 6.11 g/cm$^{-3}$). The partial density of cobalt and oxygen elements ($\rho_{Co}$ and $\rho_{O}$) are 4.49 g/cm$^{-3}$ and 1.62 g/cm$^{-3}$, respectively.
	So the attenuation lengths at 780 eV for the cobalt and oxygen elements in the Co$_3$O$_4$ are expected to be $\sim$140 nm and $\sim$700 nm, respectively. 
	But the attenuation length values at the absorption edge are likely overestimated using the Henke's table. For cobalt metal ($\rho$ = 8.9 g/cm$^{-3}$), the estimated attenuation length is $\sim$75 nm but the experimental results indicate that the attenuation length was $\sim$25 nm at the peak maximum~\cite{Chen1995prl}. 
	Thus we estimated the value within the range from 25 nm to 140 nm for the attenuation length of cobalt element at 780 eV. 
	
	The weighting of $\mu$ is proportional to the inverse of the attenuation length that weighting of $\mu$ for Co and O are estimated to be 96-83\% and 4-17\% (attenuation lengths are 25-140 and 700 nm). 
	For pure Co$_3$O$_4$, the background absorption ($\mu_B$) at 780 eV is simply the contribution of the oxygen absorption (the contributions of other edges were omitted). 
	Thus a value of $\sim$10\% of the $\mu_{max}$ is suitable estimation for the $\mu_B$. 
	
	\subsection{Used parameter values and the effective crystal field energy}
	Table~\ref{tab_Co3O4_parameter} and ~\ref{tab_Co3O4_parameter_CToff_reduce} gives the used parameter values. 
	The Slater integrals F${^\mathrm{2}_{dd}}$, F${^\mathrm{4}_{dd}}$, F${^\mathrm{2}_{pd}}$, G${^\mathrm{1}_{pd}}$, and G${^\mathrm{3}_{pd}}$ as well as the U$_{dd}$ and U$_{pd}$ are used to determine the Coulomb interaction. 
	The Slater integrals are taken to an atomic scheme, where $\sim$80\%(75\%) of the values from the Hartree-fock approximation is used for the Co$^{2+}$ (Co$^{3+}$). 
	The U$_{dd}$ and U$_{pd}$ values are set to the reference values.  
	$\zeta{_p}$ and $\zeta{_d}$ describe the spin-orbit interaction. 
	The charge transfer energy $\Delta$ and the hopping integrals V$_{e(e_g)}$/V$_{t_{2}(t_{2g})}$ mimic the energy splitting between two configurations and the electron hopping intensity from one configuration to another one. 
	
	The crystal field energy $10Dq$ identifies the energy different between the e(e$_g$) and the $t_{2}$($t_{2g}$) orbitals in the T$_d$(O$_h$) symmetry. 
	Once the ligand-to-metal charge transfer is included, the total effective crystal energy $10Dq_{\rm eff}$ is composed by two different parts: (i) ionic crystal field energy of cobalt $3d$ shell and (ii) additional contribution caused by charge transfer and exchange interaction~\cite{Wang2017jpcc}. 
	The $10Dq_{\rm eff}$ in currently work can be estimated by $^1$A$_{1g}$-$^1$T$_{1g}$ and $^4$A$_2$-$^4$T$_2$ excited states energy for the octahedral Co$^{3+}$ and the tetrahedral Co$^{2+}$ sites, which are $\sim$1.90 eV and $\sim-$0.55 eV, respectively. 
	The negative sign on the tetrahedral symmetry infers to the inverse of t$_2$ and e orbitals with respect to the octahedral symmetry. 
	In contrast, for the simulation considering the charge transfer and exchange interaction effects, the ionic crystal field energy should be further reduce to 1.15 eV and $-$0.1 eV for the octahedral Co$^{3+}$ site and tetrahedral Co$^{2+}$ site, respectively. 
	Our theoretical crystal field energy values (obtained by LDA calculation) considered only the values applied on the Co $3d$ orbitals, which means the charge transfer induced crystal field energy splitting was not involved. 
	Thus only ionic crystal field energy of cobalt $3d$ shell has been compared in the main text. 
	We note that the contraction induced by the core hole is applied to the whole valence state wave function (correspond to $10Dq_{\rm eff}$), thus we applied the $10Dq_{\rm eff}$ value of the intermediate state is reduced by $\sim$15\% in comparison with the ground state~\cite{Cramer1991jacs} (1.59 eV for the Co$^{3+}$ site and $-$0.47 for the Co$^{2+}$ site). 
	
	In the simulation, a 300 meV (FWHM) Lorentzian convoluting a 300 meV (FWHM) Gaussian was used to simulate the intrinsic broadening and the instrumental broadening of incident beam. It provides a total width 0.6 eV. For the RIXS spectra, the same incident beam width was applied. In addition to it, a 50 meV (FWHM) Lorentzian convoluting a 60 meV (FWHM) Gaussian was used for the emitted beam, which implies a total width 0.11 eV. These values are comparable to the experimental setting. Nevertheless, we note that the intrinsic broadening was fixed to a value in current simulation, where the value might depend on the selected energy. 
	
	\begin{table}[t]
		\caption{The model parameters used in the simulation (in eV), which are the Slater integral, spin-orbit coupling energies, crystal field energy, charge transfer energy, hopping integrals, U$_{dd}$ and U$_{pd}$. The \textit{i} and \textit{m} stand for the configurations of initial ground state and intermediate state, respectively. }\label{tab_Co3O4_parameter}
		\centering
		\begin{tabular}{l c c c c c c c }
			\hline
			\hline
			& F${^\mathrm{2}_{dd}}$ & F${^\mathrm{4}_{dd}}$ & F${^\mathrm{2}_{pd}}$ & G${^\mathrm{1}_{pd}}$ & G${^\mathrm{3}_{pd}}$ & $\zeta{_p}$ & $\zeta{_d}$ \\
			\hline
			\hspace{5pt}Co$^{2+}_{i}$ & 9.284  & 5.767 &   -   &   -   &   -   &   -   & 0.066 \\
			\hspace{5pt}Co$^{2+}_{m}$ & 9.917  & 6.166 & 5.808  & 4.318  & 2.455  & 9.748 & 0.066 \\
			\hline
			\hspace{5pt}Co$^{3+}_{i}$ &  9.371  & 5.859 &   -   &   -   &   -   &   -   & 0.055 \\
			\hspace{5pt}Co$^{3+}_{m}$ &  9.932 & 6.212 &  5.925  &  4.463  &  2.540 & 9.747 & 0.055 \\
			\hline
			\hline
			& $10Dq$ & $10Dq_{\rm eff}$ & $\Delta$ & V$_{e(e_g)}$ & V$_{t_{2}(t_{2g})}$ & U$_{dd}$ & U$_{pd}$ \\
			\hline
			\hspace{5pt}Co$^{2+}_{i}$ & $-$0.10 & $-$0.55 & 4.5 & 1.0 &  2.0  & 4.5  &  - \\
			\hspace{5pt}Co$^{2+}_{m}$ & $-$0.02 & $-$0.47 & 4.5 & 1.0 &  2.0  & 4.5  & 6.0 \\
			\hline
			\hspace{5pt}Co$^{3+}_{i}$ & 1.15 & 1.90 & 1.5 & 3.12 & 1.8  & 6.5  &  - \\
			\hspace{5pt}Co$^{3+}_{m}$ & 0.84 & 1.59 & 1.5 & 3.12 & 1.8  & 6.5  & 7.5 \\
			\hline
			\hline
		\end{tabular}
	\end{table}
	
	\begin{table}[t]
		\caption{The model parameters used in the case of reduced Slater integral(in eV).}\label{tab_Co3O4_parameter_CToff_reduce}
		\centering
		\begin{tabular}{l c c c c c c c }
			\hline
			\hline
			& F${^\mathrm{2}_{dd}}$ & F${^\mathrm{4}_{dd}}$ & F${^\mathrm{2}_{pd}}$ & G${^\mathrm{1}_{pd}}$ & G${^\mathrm{3}_{pd}}$ & $\zeta{_p}$ & $\zeta{_d}$ \\
			\hline
			\hspace{5pt}Co$^{2+}_{i}$ & 7.543  & 4.686 &   -   &   -   &   -   &   -   & 0.066 \\
			\hline
			\hspace{5pt}Co$^{3+}_{i}$ &  5.065  & 4.750 &   -   &   -   &   -   &   -   & 0.055 \\
			\hline
			\hline
			& $10Dq$ & $10Dq_{\rm eff}$ & $\Delta$ & V$_{e(e_g)}$ & V$_{t_{2}(t_{2g})}$ & U$_{dd}$ & U$_{pd}$ \\
			\hline
			\hspace{5pt}Co$^{2+}_{i}$ & $-$0.50 & - &  - &  - &   - &  -  &  - \\
			\hline
			\hspace{5pt}Co$^{3+}_{i}$ & 1.95 & - & - &  - &  -  &  -  &  - \\
			\hline
			\hline
		\end{tabular}
	\end{table}
	
	
	\subsection{Estimating the differential orbital covalency of an cation from the cluster model}
	Including the ligand-to-metal charge transfer effect suggests that the ground state configuration is a mixture of the $3d$ orbit and ligand hole ($\underline{L}$). 
	We calculated the weight of configurations up to two ligand holes and list them in Tale.~\ref{tab_Co3O4_weight}. 
	Then, we further estimated the cation orbital covalency of Co$^{2+}$ and Co$^{3+}$ cations using the following relation~\cite{Wang2017jpcc, DelgadoJaime2016jsr, Wasinger2003jacs}:
	\begin{equation}
	Cation~Orbital~Covalency(\gamma) = 100\%-{N}\frac{P_{\gamma}}{P_{sum}},
	\end{equation}
	where $\gamma$ stands for the state corresponding to the e(e$_g$) or t$_{2}$(t$_{2g}$) orbitals, where 100\% refers to the target orbital is dominated by the ionic configuration.  
	The coefficient ${N}$ is a renormalization factor of the number of holes in the orbit $\gamma$ out of number of holes in 3d$^n$ configuration. 
	For example, in the case of high-spin Co$^{2+}$(T$d$), there are three holes in the t$_{2}$ orbit out of three holes in 3d$^7$ configuration. Hence the renormalization factor is equal to one ($\frac{\mathrm{number~of~holes~in}~3d^n}{\mathrm{number~of~holes~in}~t_{2}} = 1$). 
	In contrast, the renormalization factor for the e orbit is meaningless because it is fully occupied (no hole exist).
	The $P_{\gamma}$ is the percentages for the configurations which accept the elections transfer from ligand to the orbital $\gamma$. 
	Note here that we only consider one electron transfer in the covalency estimation. 
	$P_{sum}$ is the percentages summation of all possible configurations involved in the hybridization, which is equal to one in current cases. 
	Thus the cation orbital covalency of t$_{2}$ orbital on the Co$^{2+}$ site and e$_g$ orbital on Co$^{3+}$ site are given as $\sim$80\% and $\sim$50\%, respectively. 
	
	\begin{table}[t]
		\caption{The weight of configurations and orbital covalency in ground state (unit in \%). Although the number of ligand holes is considered up to two in the spectral simulations, the covalency is estimated only using the configurations up to one ligand hole.}\label{tab_Co3O4_weight}
		\centering
		\begin{tabular}{l c c c c c}
			\hline
			\hline
			& $\ket{3d^n}$ & $\ket{3d^{n+1}\underline{L}^1}$ & $\ket{3d^{n+2}\underline{L}^2}$ & e(e$_{g}$) covalency & t$_{2}$(t$_{2g}$) covalency \\
			\hline
			\hspace{5pt}Co$^{2+}$($3d^7$) & 79  & 20 &   1  & 100  & 80 \\
			\hspace{5pt}Co$^{3+}$($3d^6$) &  40  & 50 & 10  & 50  & 100 \\
			\hline
			\hline
		\end{tabular}
	\end{table}
	
	\bibliography{supplement}